\documentclass[aps,amsmath,graphicx,twocolumn]{revtex4}

\usepackage{graphics}
\usepackage{graphicx}
\usepackage{amsmath}
\usepackage{amssymb}
\usepackage{epsfig}
\usepackage{epstopdf}
\usepackage{hyperref}

\begin{document}

\title{Superconductivity and Magnetism in Noncentrosymmetric RhGe}

\author{A. V. Tsvyashchenko}
\email{tsvyash@hppi.troitsk.ru}
\author{V. A. Sidorov}
\author{A. E. Petrova}
\author{L. N. Fomicheva}
\author{I. P. Zibrov}
\affiliation{L.~F.~Vereshchagin Institute for High Pressure Physics,
Russian Academy of Sciences, 142190, Moscow, Troitsk, Russia}

\author{V. E. Dmitrienko}
\affiliation{A. V. Shubnikov Institute of Crystallography, Russian Academy
of Sciences, 119333, Moscow, Russia}

\begin{abstract}
RhGe synthesized at high pressure is crystallized in noncentrosymmetric
cubic structure of the $B20$ type. Measurements of the electrical
resistivity and magnetization demonstrate a superconducting state below
$T_c \sim$ 4.5 K and a weak ferromagnetism below $T_m \sim$140 K. Specific
heat data confirm the bulk nature of superconductivity in this
ferromagnetic superconductor. The superconducting region forms a dome on
the P-T diagram with a maximum of $T_c$ near 4 GPa. {\it Ab initio}
simulations suggest that the observed weak magnetization emerges from the
pronounced spin polarization with magnetic quadrupole and toroidal moments
located at Rh and Ge sites.
\end{abstract}

%\pacs{}

\maketitle

With the discovery of the superfluid phases of $^{3}$He \cite{1,2}
understanding  of  superconductivity in terms of a condensate of Cooper
pairs, with the Cooper pairs forming due to electron-phonon interactions
\cite{3} began to change. It was found that the heavy-fermion compounds
containing $f$-elements are prime candidates for unconventional
superconductivity with complex order parameter symmetries \cite{4}. These
compounds can be put in close analogy with the heavy Fermi liquid
$^{3}$He, and  analogous interactions in that case with high probability
lead to spin-triplet, magnetically mediated, superconductivity \cite{5}.

The central issue for heavy-fermion superconductors is the question of the
coexistence of superconductivity and magnetism. First time the
superconductivity (in a limited pressure range) was discovered on the
border of itinerant-electron ferromagnetism below 1 K, in a pure  system,
UGe$_2$ \cite{6}. Superconductivity of the ferromagnet URhGe \cite{7} and
UCoGe \cite{8} observed at normal pressure was also considered in  terms
of the magnetic and spin triplet pairing as for UGe$_2$.

However for the superconducting condensate an interesting question is not
only the mechanism of pairing and the symmetry \cite{9,10}. According to
Refs. \cite{11,12} in the absence of inversion symmetry the order
parameter becomes a mixture of spin-singlet and spin-triplet components,
which leads, for instance, to the Knight shift attaining a nonzero value
at $T$ = 0. The striking examples of such noncentrosymmetric
superconductors are CePt$_3$Si \cite{13,13a} and UIr \cite{14}.

On the other hand there were unsuccessful attempts to observe
superconductivity under high pressure on the border of itinerant-electron
ferromagnetism in a noncentrosymmetric compound without $f$-elements (for
example MnSi) \cite{15,16}.

The aim of this Letter is to report on unconventional superconductivity
and long range high temperature magnetic order in the high pressure cubic
phase of RhGe (the $B20$ type \cite{17}), to evaluate parameters
characterizing the superconducting state and to discuss possible scenarios
of magnetic ordering.

The MX compounds of the transition metals (M) (e.g., Mn, Fe, Co) with
metalloids (X) (e.g., Si, Ge) which crystallize in the FeSi ($B20$)
structure type continue to attract attention in the field of solid state
physics due to the existence of rich magnetic and electronic phenomena. In
the silicides there have been found long-period helical structures
\cite{18}, quantum phase transitions and  partial order in the
high-pressure phase \cite{19,20}, vortex-like spin textures or skyrmions
\cite{21,22}, which can be controlled by an electric current, inducing a
topological Hall effect \cite{23,24,25}.

The cubic high pressure phase of MnGe \cite{26} shows the highest magnetic
moment among the $B20$ metals \cite{27,28} which around 6 GPa transforms
from a high-spin to a low-spin state \cite{29}. Germanides with the $B20$
structure have stronger ferromagnetic properties  than the silicides
\cite{30,31}, as required to exploit their chiral magnetism.

Polycrystalline samples of RhGe cubic phase  were synthesized at 8 GPa in
the toroidal high-pressure apparatus \cite{32} by melting reaction between
Rh and Ge. The purity was 99.99\% for Rh and 99.999 \% for Ge. The pellets
of well-mixed powdered constituents were placed in rocksalt pipe ampoules
and then directly electrically heated to 1700 K. Then the samples were
quenched to room temperature before the applied pressure was released.
Crystal structure was determined from x-ray data (Cu K$\alpha_{1}$
radiation, Guinier camera - G670, Huber) and found to be simple cubic, the
space group $P2_{1}3$ (No. 198) with $a$ =4.85954(2) \AA~($V$ = 114.758(1)
\AA$^{3}$), isotypic with the crystal structure of FeSi ($B20$) without
the inversion symmetry. The $B20$ structure has 8 atoms per unit cell.
Both Rh and Ge are located at the Wyckoff positions (4$a$) with
coordinates ($u$, $u$, $u$), ($u + 0.5$, $0.5 - u$, $- u$), ($-u$, $0.5 +
u$, $0.5 - u$), and ($0.5 - u$, $- u$, $0.5 + u$) where $u$(Rh) =
0.12809(11) and $u$(Ge) = 0.83368(13). Rietveld refinements (GSAS
\cite{33,34}) revealed phase purity of the polycrystalline material used
for bulk property measurements. The sites occupancy of Rh and Ge was also
determined from the Rietveld analysis (Rh/Ge $\sim$ 1/0.986 for this
sample). More details can be found in Supplementary Materials \cite{35}.

\begin{figure}
\includegraphics[width=8cm]{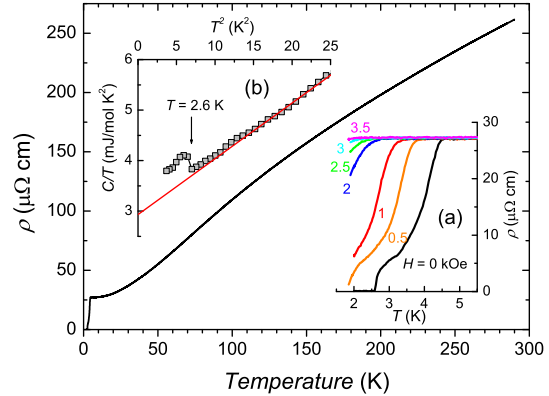}
\caption{\label{fig1} Temperature dependence of the electrical resistivity
of RhGe$_{0.986}$. Inset (a) - temperature dependences of resistivity in
different magnetic field around the superconducting transition. Inset (b)
- temperature dependence of the specific heat (in the form $C/T$ vs
$T^{2}$) of RhGe around the superconducting transition.}
\end{figure}

\begin{figure}
\includegraphics[width=8cm]{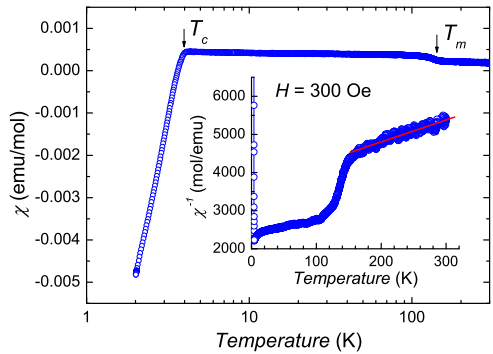}
\caption{\label{fig2} Temperature dependence of the magnetic
susceptibility of RhGe$_{0.986}$.  The inset shows temperature dependence
of the inverse susceptibility. The bold red line in the inset is a linear
fit of data above the magnetic transition.}
\end{figure}

\begin{figure}
\includegraphics[width=8cm]{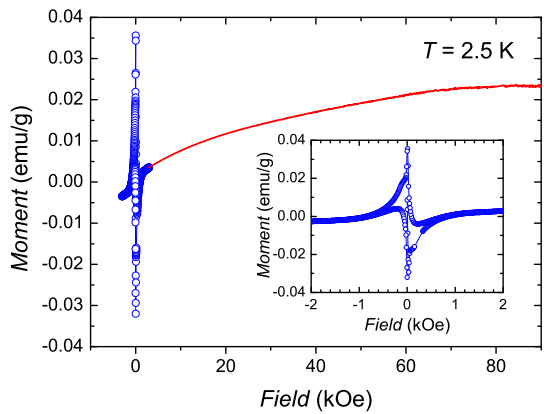}
\caption{\label{fig3} Field dependence of the magnetization of
RhGe$_{0.986}$ at 2.5 K. The inset shows the enlarged plot of
magnetization at low fields.}
\end{figure}

Basic physical properties of this polycrystal are displayed in Figs.
\ref{fig1}, \ref{fig2}, \ref{fig3}. The electrical resistivity and
specific heat were measured with the Quantum Design PPMS instrument.
Magnetic properties were measured with VSM inserted in PPMS. Temperature
dependence of the electrical resistivity $\rho(T)$ of RhGe$_{0.986}$ (Fig.
\ref{fig1}) is of metallic type with gradual saturation below 10 K. In the
range 4.5-30 K $\rho(T)$ follows the dependence $\rho(T) = \rho_{0} +
AT^{3}$. This type of dependence was found in a number of transition metal
compounds and is explained by phonon assisted interband {\it s-d}
scattering \cite{36}. Below 10 K, a $T^{2}$ dependence of $\rho(T)$
typical for the Fermi liquid was found  in FeGe and MnGe \cite{37}. At 4.5
K resistivity starts to drop and finally goes to zero at 2.6 K. Thus RhGe
is the first superconductor in the series of MX compounds with the $B20$
structure lacking inversion symmetry. Earlier superconductivity with
$T_{c}$ = 0.96 K was reported for an ambient pressure phase of RhGe having
the orthorhombic ($B31$) MnP-type structure \cite{38}. Nevertheless for
RhSi ($B20$) the superconductivity above $T_{n}$ = 0.35 K was not detected
\cite{38}. The inset (a) in Fig. \ref{fig1} shows $\rho(T)$ around the
superconducting transition in different magnetic fields up to 3.5 kOe for
RhGe$_{0.986}$. We estimated $dH_{c2}/dT$ at $T_{c}$ as $-0.65$ kOe/K. The
shape of $\rho(T)$ near the superconducting transition may indicate  the
presence of two superconducting phases.

The bulk nature of the transition is confirmed by the specific heat
measurements (inset (b) in Fig. \ref{fig1}). The bold straight line in the
plot $C/T$ vs $T^{2}$ is the best fit of data between 4 and 7 K and
represents the contribution from phonons and electrons in the normal
state. The electronic specific heat coefficient $\gamma$ = 2.9
mJ/(mol$\cdot$K$^{2})$ allows one to classify RhGe as a normal metal with
weak electronic correlations. For magnetic MnGe, FeGe the coefficients
$\gamma$ are 16 and 9 mJ/(mol$\cdot$K$^{2})$, respectively and for
nonmagnetic CoGe $\gamma$ $\sim$ 0 mJ/(mol$\cdot$K$^{2})$ \cite{37}. The
jump of the specific heat takes place at 2.6 K where resistivity goes to
zero. Also small deviation between the bold line and experimental data may
indicate on the contribution of superconducting phase to the specific heat
below 4 K. The jump of the specific heat at the superconducting transition
is small $\Delta C/\gamma T_{c} \sim$ 0.16, and well below the BCS value
$\Delta C/\gamma T_{c}$ = 1.43. A reduced value of the specific heat jump
at $T_{c}$ may arise if only some parts of the Fermi surface have non-zero
superconducting gaps, while the others remain gapless. Similar small jump
$\Delta C/\gamma T_{c}$ = 0.25 was found in the noncentrosymmetric
antiferromagnetic superconductor CePt$_{3}$Si \cite{13} and in the
ferromagnetic superconductor UGe$_{2}$ with the maximum of  $\Delta
C/\gamma T_{c} \sim$ 0.29 and the maximum of $T_{c}$ = 0.6 K  reached at a
critical pressure 1.22 GPa \cite{39}.

Notice that according to the measurements of magnetic susceptibility
$\chi(T)$ in applied field $H$ = 300 Oe (Fig. \ref{fig2}) RhGe becomes
weakly ferromagnetic below $T_{m} \sim$ 140 K and superconducting below
4.1 K. Measurements at 2.5 K (Fig. \ref{fig3}) show the picture of
magnetization  typical for a ferromagnetic superconductor,  demonstrating
the coexistence of superconductivity and ferromagnetism  at 2.5 K.
Magnetic moment of RhGe$_{0.986}$ is saturated at the level of 0.023 emu/g
($\sim$0.0007 $\mu_{B}$/f.u.) in magnetic fields above 70 kOe. This
extremely small value of saturation moment should be compared with an
effective magnetic moment in the paramagnetic state $\mu_{eff}$ = 1.18
$\mu_{B}$/f.u. found from the slope of a dependence $\chi^{-1}(T)$ (inset
in Fig. \ref{fig2}). Thus, RhGe is a weak itinerant ferromagnet, but with
a rather strong magnetic exchange ($T_{m} \sim$ 140 K).

We have measured the low field ($\sim$1 Oe) magnetic ac-susceptibility to
estimate the superconducting volume fraction of RhGe compared to typical
BCS-type superconductor (Pb). Sample of RhGe and a piece of Pb with
external dimensions close to those of RhGe sample were cooled down to 1.7
K in a home-made coil system. We tested RhGe$_{0.986}$ and RhGe$_{0.989}$
samples. For both samples the magnitude of diamagnetic signal below
$T_{c}$ was similar to that produced by Pb at 7.2 K. It was also found
that the appearance of superconductivity is sensitive to sample
composition and is favorable for samples close to stoichiometric RhGe. The
RhGe$_{0.98}$ sample though has not displayed superconducting properties.

\begin{figure}
\includegraphics[width=8cm]{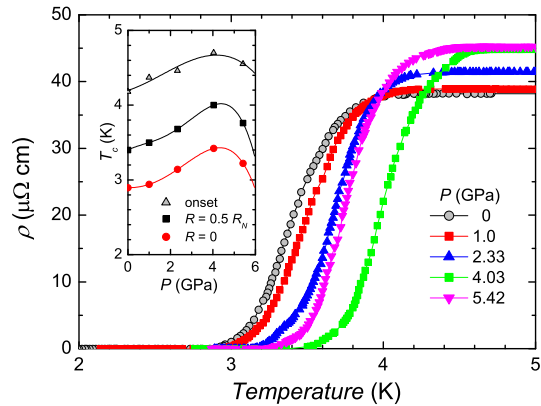}
\caption{\label{fig4} Temperature dependence of the electrical resistivity
of RhGe$_{0.989}$ near the superconducting transition at different
pressures. The inset shows the P-T diagram of superconductivity in RhGe.}
\end{figure}

Pressure effect on superconductivity was studied for another
superconducting sample of RhGe$_{0.989}$. Temperature dependences of the
electrical resistivity $\rho(T)$ of RhGe$_{0.989}$ around the
superconducting transition at different hydrostatic pressures are shown in
Fig. \ref{fig4}. Clamped toroid-type pressure cell with a liquid pressure
medium \cite{40} was used for this experiment. The width of the
superconducting transition does not depend on hydrostatic pressure. The
superconducting temperature $T_{c}$ increases first at high pressure at a
rate $dT_{c}/dP$ = 0.15 K/GPa but above 4 GPa $T_{c}$ begins to decrease.

Structural, electronic and magnetic properties of the RhGe crystals of the
$B20$ and MnP types were quantitatively evaluated with the Quantum
ESPRESSO package developed \cite{41} for the density-functional-theory
(DFT) computations. The exchange-correlation functional with the
generalized gradient approximation (GGA) of the Perdew-Burke-Ernzerhof
type was chosen.

The scalar relativistic pseudopotentials were used for structure
optimization (all the pseudopotentials were downloaded from the Quantum
ESPRESSO data base \cite{42}). To speed up the calculations, we used the
symmetrized $k$-points: $6\times6\times6$ mesh for cubic $B20$ structure
and $6\times10\times6$ for orthorhombic MnP-type structure. The plane-wave
basis for wavefunctions had the cut-off energy of 40 Ry and the density
cut-off energy was 440 Ry. The calculated energy-vs-volume dependence for
RhGe is discussed in the Supplementary Materials. The $B20$ structure is
more dense and it becomes energetically favorable for pressure above circa
8 GPa  (at $T$ = 0).

Fully relativistic GGA pseudopotentials were used for searching possible
non-collinear magnetic structures with the spin-orbit interaction. The
magnetic effects have been found to be very subtle in the case of the
so-called ultra-soft potentials and we will not present correspondent
simulations. It has been found that the final results do not change much
for the $k\times k\times k$ between $8\times8\times8$ and
$10\times10\times10$ and for the wavefunction cut-off energy between 40 Ry
and 80 Ry. The pictures of the magnetization distribution $\mathbf
M(\mathbf r)$ presented below have been obtained for the $9\times9\times9$
mesh and the 40 Ry cut-off energy.

\begin{figure}
\includegraphics[width=8cm]{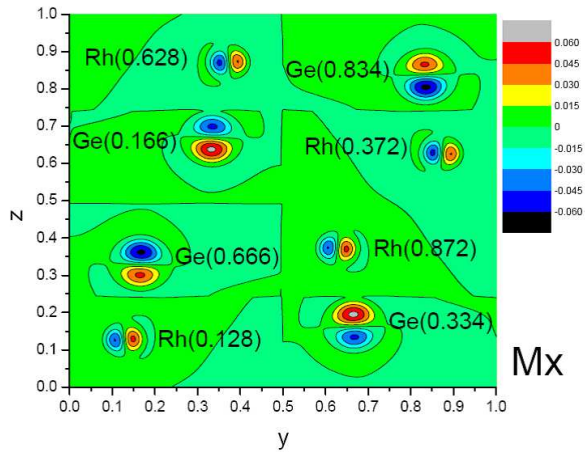}
\caption{\label{fig5} Magnetic ordering in the RhGe unit cell: the space
distribution of the $M_{x}$-magnetization around Rh and Ge atoms viewing
along the $x$-axis (Quantum Espresso simulations). Similar patterns for
the $M_{y}$-  and $M_{z}$-components can be found in Supplementary
Materials \cite{35}. The figure is composed from 8 patches showing the
magnetization in the $yz$-planes passing through the centers of
corresponding atoms at 8 different $x$-levels; the atomic symbols and the
$x$-levels are indicated in the figure. The boundaries between the patches
are the straight lines: $y$=0.5, $z$=0.25, $z$=0.5, and $z$=0.75. The
color scale palette is in arbitrary units. }
\end{figure}

A typical example of the relaxed magnetization distribution inside a unit
cell is shown in Fig. \ref{fig5}. The most interesting features are the
small neighboring regions with plus and minus magnetizations along $x$-,
$y$-, and $z$-directions. The shapes of the regions are typical of the $p$
electronic states and their sizes scale with $Z^{-1}$ as expected
($Z_{Rh}/Z_{Ge}\approx1.4$). The orientations of the regions are different
for different atoms  in accordance with the orthorhombic symmetry
$P2_{1}2_{1}2_{1}$ (instead of cubic $P2_{1}3$ because of the uniaxial
initial magnetization). The calculated total magnetization is rather
small and directed along $z$, $\lvert\langle \mathbf M(\mathbf r) \rangle
\rvert \approx 0.0077\ \mu_{B}$, whereas the average absolute
magnetization $\langle \lvert\mathbf M(\mathbf r)\rvert \rangle \approx
1.0\ \mu_{B}$. We see that $\langle \lvert \mathbf M(\mathbf r)\rvert
\rangle \gg \lvert \langle \mathbf M(\mathbf r) \rangle \rvert$, which
means mainly `antiferromagnetic' distribution of the magnetization. The
absolute magnetization can be better characterized by the tensor of
magnetization directions $D_{ik} = \langle M_{i}(\mathbf r)M_{k}(\mathbf
r)  \rangle$. However for the whole unit cell this tensor is almost
spherically symmetric and does not tell much about the details of $\mathbf
M(\mathbf r)$.

For quantitative characterization of atomic magnetization we calculated
different magnetic quantities for spheres around atoms. The radius of the
spheres is chosen equal to 0.64 \AA~or 0.13 in the crystallographic units;
see Fig. \ref{fig5}, which for the Rh atom roughly corresponds to the
distance from the atom center to the unit cell boundary at the down-left
corner. Obviously such spheres include the most interesting features of
the magnetization distribution. The total volume of 8 spheres is only
7.8\% of the unit cell. In the following we present results for Rh and Ge
atoms at standard positions (at $u$, $u$, $u$). Data for the other sites
can be obtained by corresponding symmetry transformations.

We have found that the atomic magnetization averaged over the sphere  is $
\langle \mathbf M( \mathbf r) \rangle _{sph} = (170,55,20)\cdot 10^{-4}
\mu_{B}$ for Rh and $(2,-4,-3)\cdot 10^{-4}\mu_{B}$ for Ge. It is
interesting that for Rh the average atomic moment is directed mainly in
$x$-direction whereas for Ge the $x$, $y$, and $z$ components are
comparable. However after summation over four RhGe units the $x$ and $y$
components almost vanish and only $z$ magnetization survives providing
almost 90\% of the total ferromagnetic magnetization. In contrast, the
absolute magnetization inside the spheres gives only 40\% of the absolute
magnetization, $\langle \lvert \mathbf M (\mathbf r) \rvert \rangle _{sph}
= 0.04 \mu_{B}$ per Rh and 0.06 $\mu_{B}$ per  Ge. The remaining 60\% is
an itinerant magnetization distributed in 92\% of the unit cell volume.

The spatial distribution of magnetization inside the atomic spheres  can
be characterized by a non-symmetric tensor $Q_{jk} = \langle r_{j}
M^{'}_{k} \rangle_{sph}$ where $\mathbf M^{'}(\mathbf r) = \mathbf
M(\mathbf r) - \langle \mathbf M(\mathbf r) \rangle _{sph}$. This tensor
violates both $T$- and $P$-invariance \cite{43} and can appear, for
instance, if a $p$-state is admixed to an $s$-state. The symmetric part
$(Q_{jk}+Q_{kj})/2$ is the magnetic quadrupole moment of the atom and the
antisymmetric part $(Q_{jk}-Q_{kj})/2$ is equivalent to the vector of
atomic toroidal moment $\mathbf T = \langle \mathbf r \times \mathbf M^{'}
(\mathbf r) \rangle _{sph}$. It is interesting that there is a non-zero
scalar part $Q_{jj}$ arising  from the spin hedgehog pattern around each
atom \cite{44}: $Q_{jj}$ is of about $1.5\cdot 10^{-4}\ \mu_{B}$ for Rh
(plus for two atoms and minus for another two) and  $1.0\cdot 10^{-4}\
\mu_{B}$ for Ge. Thus each atom in RhGe has a complicated magnetic pattern
and can be considered as an atomic-size Skyrmion. The calculated
toroidal moment for Rh is of about $2.6 \cdot 10^{-4} \mu_{B}$  whereas
for Ge it is only $0.2 \cdot 10^{-4} \mu_{B}$. In these calculations
vector $\mathbf r$ is in dimensionless crystallographic units (and
therefore both $Q_{jk}$ and $\mathbf T$ have the dimension of $\mu_{B}$).

Our calculations of possible magnetic ordering in RhGe with the $B20$
structure are mainly illustrative rather that quantitative. However, they
allow us to understand the physical picture behind them. Both Rh and Ge
atoms possess `antiferromagnetically' distributed local spin densities
with quadrupole and toroidal magnetic moments regularly ordered inside the
unit cell. The plus and minus magnetization directions are slightly canted
(probably owing to the spin-orbit Dzyaloshinskii-Moriya interactions).
Since these cants governed by the crystal symmetry are regular, small
ferromagnetic moments emerge both for each atom and for the unit cell as a
whole. Different orientations of $\mathbf M(\mathbf r)$ for different
atoms can be observed in magnetic neutron or x-ray diffraction via $h00$,
$0k0$, $00l$ forbidden reflections with odd $h$, $k$ or $l$.

It is possible what both the low-temperature superconducting state and
magnetic ordering may be explained by mixing $s$- and $p$-wave state.
Since the net ferromagnetic moments are weak, they should not be
devastating for superconductivity. From the other side, the possible role
of toroidal moments in superconductivity is discussed for many years
\cite{45, 46} and our findings could bring a new dimension to those
discussions.

Of course, the suggested picture of magnetic ordering is oversimplified
because it does not include possible helimagnetic spiraling typical of the
noncentrosymmetric $B20$ structures. Perhaps more sophisticated
pseudopotentials and the Hubbard $U$ interaction should be used for a more
quantitative description of magnetism in RhGe.

In conclusion, we have discovered an unusual co-existence of
superconductivity and magnetism in the noncentrosymmetric RhGe crystal. An
importance of the spin-orbit interactions for non-collinear magnetization
distributions characterized by magnetic quadrupole, toroidal, and
spin-hedgehog-like patterns is demonstrated with {\it ab initio}
simulations. Our findings provide a basis for the further studies of
noncentrosymmetric magnetic superconductors.

The authors are grateful to S. M. Stishov, for support of this work and
for helpful discussion. A. V. T. thanks  A.V. Nikolaev and D.A. Salamatin
for helpful discussion. The work was supported by the Russian Foundation
for Basic Research (grants No. 14-02-00001), the Russian Science
Foundation (grant RSF-14-22-00093) and by special programs of the
Department of Physical Science, Russian Academy of Sciences.

\widetext
\clearpage
\begin{center}
\textbf{\large Supplemental materials to the manuscript}
\end{center}
\setcounter{equation}{0}
\makeatletter
     \@addtoreset{figure}{section}
\makeatother
\setcounter{page}{1}

\renewcommand{\theequation}{S\arabic{equation}}
\renewcommand{\thefigure}{S\arabic{figure}}
\renewcommand{\bibnumfmt}[1]{[S#1]}
\renewcommand{\citenumfont}[1]{S#1}

\begin{section}{Experimental and simulation details}
The new polymorph of RhGe with FeSi ($B20$) structure was obtained by
melting reaction of the constituent materials at 8 GPa and 1700 K. Its
powder XRD pattern was collected on HUBER (G670) diffractometer with Cu
K$\alpha_{1}$ radiation in the transmission mode, at $2 \theta$ step
0.005$^{\circ}$ in the angular range from 4 to 95 degrees at room
temperature and normal pressure. The crystal structure was refined
by Rietveld full-profile analysis of XRD pattern, using \cite{1s,2s}
programs. It crystallizes in simple cubic, space group $P2_{1}3$ (No. 198)
with $a$=4.85954(2) \AA~(see Fig. \ref{fig1s}). The unit cell  volume ($V =
114.758(1)$ \AA$^{3}$) of B20 cubic phase of RhGe is 4.6\% less than the
unit cell volume ($V = 120.042$ \AA$^{3}$) of normal pressure phase with the
orthorhombic ($B31$) MnP-type structure \cite{3s}.

\begin{figure}[h]
\includegraphics[width=8.6cm]{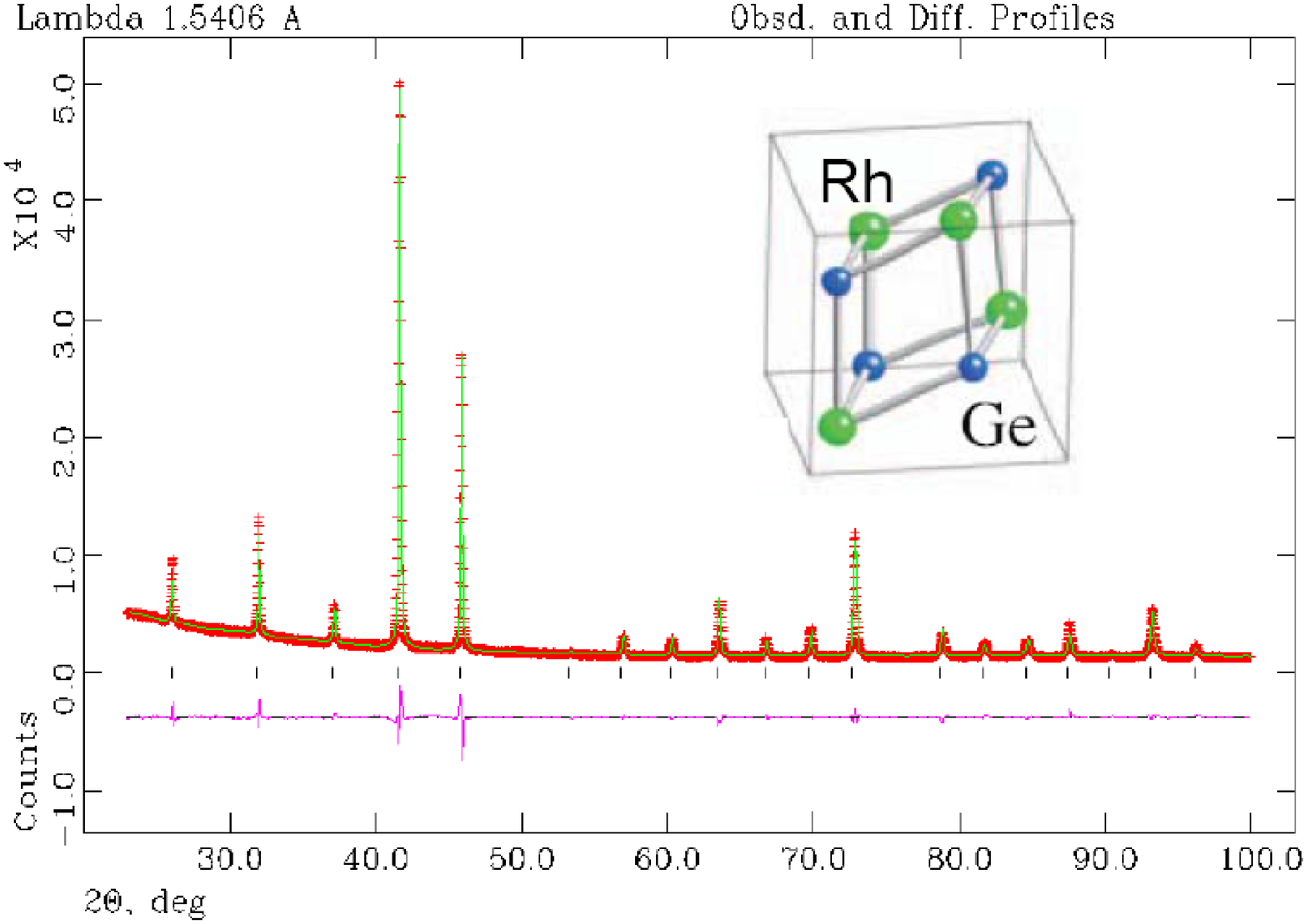}
\caption{\label{fig1s}X-ray Rietveld refinement of RhGe$_{0.986}$ ($R_{F}
= 0.0188$, $R_{P} = 0.0260$, $R_{WP} = 0.0418$). The observed (+),
calculated (solid line) and difference between observed and calculated
(bottom curve) powder diffraction profiles. The positions of all allowed
Bragg reflections are indicated by the vertical tick marks. Insert: the
crystal structure of $B20$-type RhGe.}
\end{figure}

Structural, electronic and magnetic properties of the RhGe crystals of B20
and MnP types were quantitatively evaluated with the Quantum ESPRESSO
package \cite{4s}. The scalar relativistic pseudopotentials
\verb|Rh.pbesol-spn-kjpaw_psl.0.2.3.UPF| and
\verb|Ge.pbesol-dn-kjpaw_psl.0.2.2.UPF| were used for structure
optimization (all the pseupotentials used in this work were taken from the
Quantum ESPRESSO data base \cite{5s}). The calculated energy-vs-volume
dependence for RhGe is shown in Fig. \ref{fig2s}. The $B20$ structure is
more dense and it becomes energetically favorable for pressure above circa
8 GPa (at $T$ = 0). The atomic parameters $u$ demonstrate rather strong
pressure dependence and the Rh sublattice is more changeable than the Ge
sublattice (see the insertion in Fig. \ref{fig2s}). This may be
important for the electron-phonon interaction responsible for
superconductivity. For very high pressures, the atomic parameters tend to
their ideal values $u$(Rh) = 1-$ u$(Ge) = $1/(4\tau) \approx 0.1545085$
($\tau = (1+\sqrt{5})/2$ is the golden mean) which correspond to the
perfect structure of a crystalline approximant of icosahedral
quasicrystals \cite{6s}. From the physical point of view those values
provide the densest packing of equal spheres in the B20 structure
\cite{7s}. An additional finding from the DFT simulations is that the
$B20$ phase remains stable relative to the transition into the $B2$
primitive cubic phase (the $Pm3m$ space group) even at very high pressure.

Fully relativistic GGA pseudopotentials
\verb|Rh.rel-pbesol-spn-kjpaw_psl.0.2.3.UPF| and
\verb|Ge.rel-pbesol-dn-kjpaw_psl.0.2.2.UPF| were used for searching
possible non-collinear magnetic structures with the spin-orbit
interaction. We have also tried the so-called ultra-soft (US) potentials
\verb|Rh.rel-pbesol-spn-rrkjus_psl.0.2.3.UPF| and
\verb|Ge.rel-pbesol-dn-rrkjus_psl.0.2.2.UPF| which provide more quick
convergence of the DFT process. However the magnetic effects have been
found to be very subtle in the case of the US potentials and we will not
present here the results obtained with them.

\begin{figure}[h]
\includegraphics[width=8.6cm]{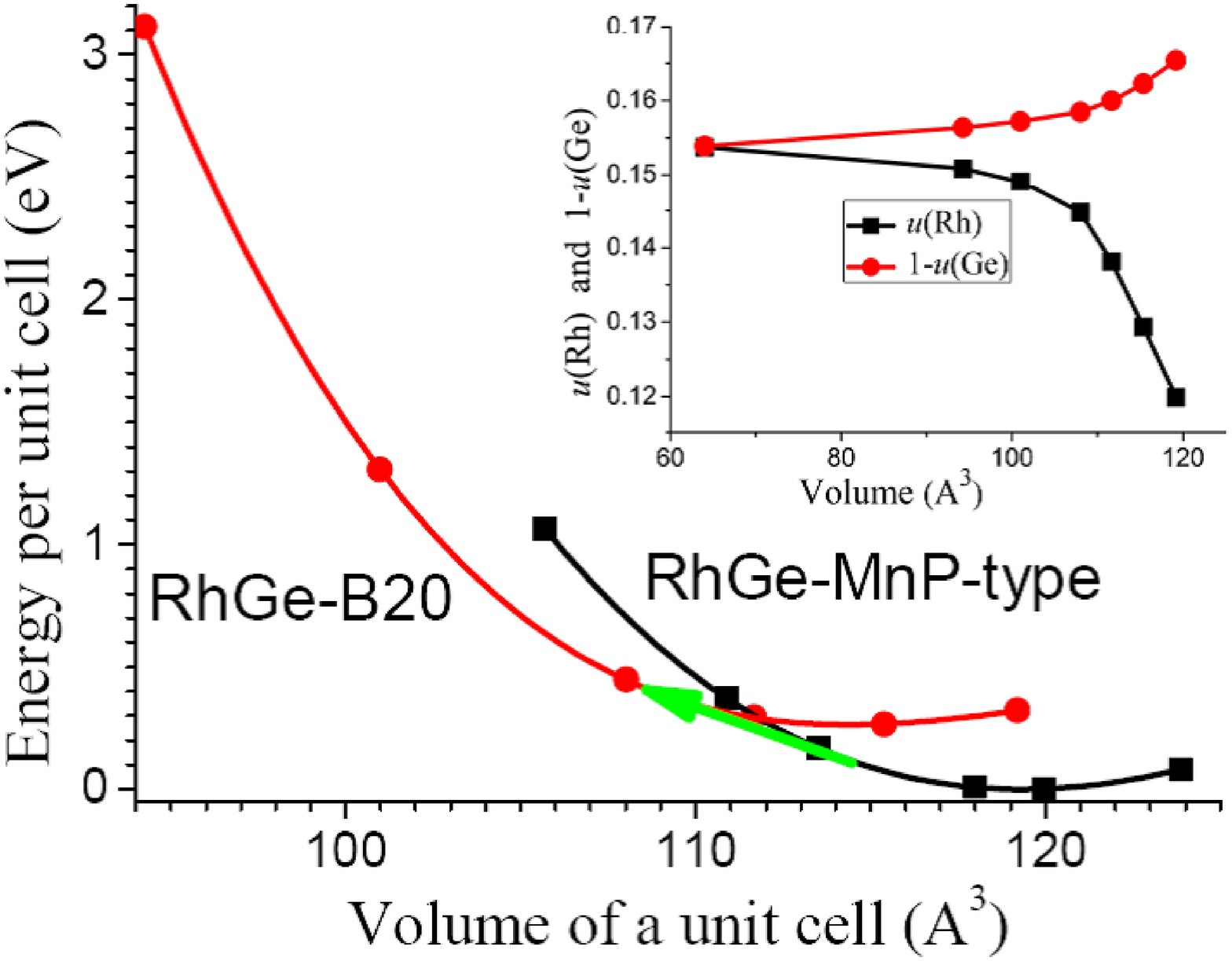}
\caption{\label{fig2s}Calculated energy vs the unit cell volume for
metastable  ($B20$) and stable (MnP-type) phases of   RhGe. The common
tangent line (shown as an arrow) determines the transition pressure (about
8 GPa) and the volume change at the transition point (about 6\%). Zero
energy corresponds to the MnP-type phase at zero pressure. Insert: changes
of the Rh and Ge atomic parameters $u$ in the $B20$ phase.}
\end{figure}
\begin{figure}[h]
\includegraphics[width=17.2cm]{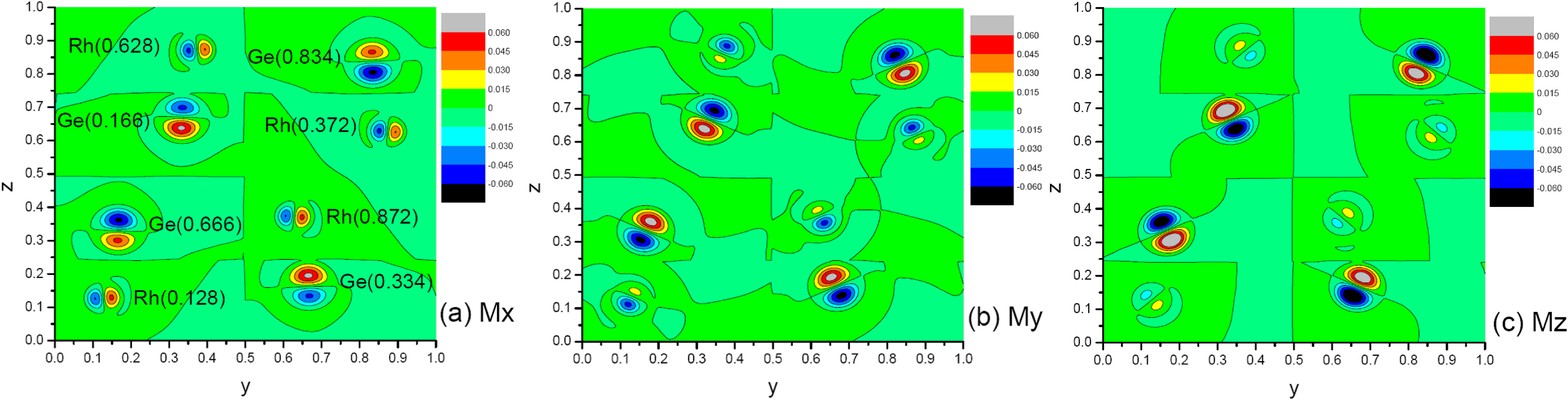}
\caption{\label{fig3s}Magnetic moment distribution inside the RhGe unit
cell viewing along the $x$-axis (Quantum Espresso simulations). Space
distributions of (a) $M_{x}$-, (b) $M_{y}$-, and (c) $M_{z}$-components of
magnetic moments are shown in the $yz$-planes passing through the centers
of corresponding atoms. Each figure is composed from 8 patches showing the
magnetization distributions around the corresponding atoms at 8 different
$x$-levels; the atomic symbols and the $x$-levels of those patches are
indicated only for (a). The boundaries between the patches are the
straight lines: $y=0.5$, $z=0.25$, $z=0.5$, and $z=0.75$. The color scale
palette is in arbitrary units.}
\end{figure}

It is expected from the cubic symmetry of the $B20$ phase that the
preferable orientation of the total magnetization $\langle \mathbf
M(\mathbf r) \rangle$ (averaged over the unit cell) should be either
\{001\} or \{111\}. Both cases have been calculated using initial
magnetizations of Rh and Ge atoms with the corresponding symmetries and
the \{001\} orientation is found to be slightly more preferable. The
magnitudes of the initial magnetizations are chosen to be of the order of
$\mu_{B}$ or less, and directed along the \{001\} axis with the same or
opposite signs for Rh and Ge atoms; in all cases they converge to similar
final values. It should be emphasized that even if the value of initial
magnetization is very small the magnetization $\mathbf M(\mathbf r)$ does
not converge to zero. Instead, the average absolute magnetization $\langle
\lvert \mathbf M(\mathbf r)\rvert \rangle$ grows during the
self-consistent energy minimization process so that the final
magnetization distribution has cubic symmetry with zero average
magnetization $\langle \mathbf M(\mathbf r) \rangle$. However this cubic
distribution has a slightly higher energy than the distribution resulting
from the uniaxial initial magnetization and we will not present its
structure here. In all cases, the self-consistent process started from a
superposition of atomic orbitals plus a superimposed `randomization' of
atomic orbitals suggested by Quantum Espresso.

A typical example of the relaxed magnetization distribution inside a unit
cell is shown in Fig. \ref{fig3s}. Both Rh and Ge atoms demonstrate
quickly alternated patterns of inhomogeneous magnetization so that their
average magnetization is rather small. The tensor of magnetization
directions $D_{ik} = \langle M_{i}(\mathbf r)M_{k}(\mathbf r) \rangle
_{sph}$ better characterizes distribution of the magnetization near atoms
than $\langle \lvert \mathbf M(\mathbf r) \rvert \rangle _{sph}$ because
describes also the preferable direction of  the magnetization. For Rh atom
\[D(\textrm{Rh}) = \begin{pmatrix}
1.5 & 0.5 & 0.3 \\
0.5 & 2.5 & -0.1 \\
0.3 & -0.1 & 0.2
\end{pmatrix}\cdot 10^{-8}\ \mu_B^2 \]
The tensor components of $D_{ik}$(Rh) show that the Rh moments are
preferably oriented in the $xy$ plane because $D_{xx}$(Rh) and
$D_{yy}$(Rh) are larger than all other tensor components.

In contrast, for Ge all the tensor components are of the same order of
magnitude
\[D(\textrm{Ge}) = \begin{pmatrix}
7.6 & -5.2 & -3.6 \\
-5.2 & 6.5 & 4.7 \\
-3.6 & 4.7 & 11.5
\end{pmatrix}\cdot 10^{-8}\ \mu_B^2 \]
and they are larger than for Rh; this correlates with stronger absolute
magnetization of Ge atoms clearly visible in Fig. \ref{fig3s}.

\end{section}

\end{document}